# A relativistic model for Quantum Plasmas

Priya Mishra and Punit Kumar[*]

*Department of Physics, University of Lucknow- 226007*

[*]*kumar_punit@lkouniv.ac.in.*

**Abstract -** A new model to study the dynamics of relativistic quantum plasmas using the quantum electrodynamical (QED) approach has been constructed to analyze the quantum effects, relativistic corrections, and electromagnetic interactions. Considering the covariant Lagrangian function and Euler-Lagrange equation, the equations of motion have been established describing the interaction of strong electromagnetic waves in plasma. These equations of motion constitute a model for the propagation of relativistic laser pulse through high density quantum plasma. Our model specifically takes the effects of four spin and four velocity into account during the interaction process. This model is applicable to high density plasmas in all ranges of electromagnetic fields which includes astrophysical environments, high power laser plasma interactions, etc.

**Introduction** - Plasma has been hot topic for research due to its wide range of technological applications [1]. Further, as exploration revealed high density astrophysical regime, metallic plasmas and high energy requirement in dense plasmas for fusion technologies, the focus has shifted to the regime of relativistic quantum plasmas [2]. Quantum plasmas are characterized by the dynamics of collective plasma behaviour and quantum effects originating due to wave function overlap. These quantum effects include quantum tunneling, quantum diffraction and spin effects. Expanding upon the principles of high density plasmas, it represents a state of matter where the particle density is so high that the interaction between particles is dominated by quantum and collective plasma effects. Such plasmas are typically found in extreme astrophysical systems like the white dwarfs, neutron stars [3, 4], as well as in advanced laboratory experiments, and in inertial confinement fusion [5]. The degeneracy plays an important role in this regime, as it occurs at high densities and low temperature, where the quantum nature of particles cannot be ignored. At even higher densities and energy, the effects of relativity become significant giving rise to the study of relativistic quantum plasma, where both quantum and relativistic corrections are essential to describe the behaviour of particles moving with the speed comparable to the speed of light [6]. The relativistic quantum plasmas have their potential to enhance our understanding of astrophysical phenomena [7], as they optimize the next generation laser plasma interaction experiments, develop compact particle accelerators and controlled nuclear fusion [8] etc. To study the dynamics of quantum plasma there are various approaches, both in relativistic and non relativistic regimes [9]. Considering the existing approaches like hydrodynamical approach [10,11], kinetic approach [12,13] and spin resolved approaches [14,15] which incorporate quantum corrections such as Bohm potential, Fermi pressure, spin dynamics, magnetization effects, investigating high intensity pair- plasma interaction and pair production, but faced limitations in terms of accuracy, incorporating relativistic effects, and also fail to describe the higher order quantum effects and non- linear interactions. Building upon the foundational theoretical work of Tsytovich [16], Jancovici [17], and Lindhard [18], numerous studies have significantly advanced our understanding of relativistic quantum plasmas. Early QED approaches investigated phenomena such as quantum fluctuations in subcritical fields, wave propagation in magnetized plasmas, and low-frequency waves in electron-positron plasmas, considering photon-photon scattering. Although these investigations, including those utilizing the Wigner function approach [19], have provided valuable insights, higher order quantum effects remain to be fully explained. All the previously mentioned works have drawbacks that the assumptions of taking perturbative analysis to low orders become less accurate when dealing with strong e. m. fields, and also some terms in QED have been ignored. Although these studies considered relativistic plasmas, they have not fully incorporated all relativistic corrections, having ignored the coupling of quantum effects at higher order. To address these gaps, we have chosen to construct a complete and consistent treatment of relativistic effects, by using all the elements of QED to develop our model over the other approaches. This provides a more accurate description of quantum effects, higher order relativistic corrections and electromagnetic interaction with particle and spin by considering the interaction of strong e. m. waves where higher order quantum effects are important. Phenomena like particles creation, annihilation and quantum fluctuations which accounts for spin, e. m. interaction and radiative effects will become easier to deal with relativistic and quantum effects make it ideal for studying high intensity fields, ultra dense plasma high energy collision and astrophysical systems.

**Basic equations** - In the present letter, we have developed a model for propagation of relativistic laser pulse whose field amplitudes are functions of space and time, through high density quantum plasma by adopting the QED approach. Dirac's equation in tensor form has been employed to define the least action principle [20,21] incorporating all QED, quantum, and relativistic effects to formulate the equations of motion, using the time slicing technique [22]. This introduces a novel approach by allowing the laser strength applicable to lasers of all strengths, including those exceeding the Schwinger limit [23-25]. Using the covariant Lagrangian and Euler-Lagrange equations, we derive a set of equations describing the interaction of strong electromagnetic waves with spin in plasma. The analysis highlights the influence of relativistic effects through the relativistic factor, the time component of the four-spin, and coupling with the velocity field. We employ a range of Lagrangian densities to derive the equations of motion,

$$L = L_{HEC} + L_O + L_F + L_S + L_{F.P} + L_{B.P} \tag{1}$$

where, $L_{HEC} = \dfrac{\varepsilon_0^2 K}{16}\left[4\left(F_{\mu\upsilon}F^{\mu\upsilon}\right)^2 + 7\left(F_{\mu\upsilon}\hat{F}^{\mu\upsilon}\right)^2\right]$ is the Lagrangian density function of Heisenberg-Euler correction due to first order strong field QED effects [26], $\varepsilon_0$, $K$ are the free space permittivity and coupling parameter respectively. $F^{\mu\upsilon}$ is the electromagnetic field tensor and $\hat{F}^{\mu\upsilon}$ is its dual e. m. field strength tensor, $\mu$ and $\upsilon$ are commonly used as Greek indices that range over the four spacetime dimensions (0, 1, 2, 3), where 0 typically denotes the time component and 1, 2, 3 denote the spatial components. $K = \dfrac{2\alpha_{em}^{'2}\hbar^3}{45m^4c^5}$, $\alpha_{em}' = \dfrac{e^2}{4\pi\hbar c\varepsilon_0}$ is the fine structure constant, where $e$ is the electron charge.

$L_O = \dfrac{-mc^2}{\gamma}$, $L_O$ being the Lagrangian density function of free electron [27], $m, c, \gamma$ are the rest mass of electron, the light speed of electron in vacuum and the relativistic factor of electron respectively. $L_F = \dfrac{-e}{\gamma c}u_\alpha A^\alpha$, $L_F$ is the ordinary electron's e. m. field interaction Lagrangian density [27], $u_\alpha$ is the four-velocity of electron and $A^\alpha$ is the four-vector potential. $L_S = \sigma^\mu \left( \dot{\sigma}^\mu - \alpha F^{\mu\nu} \sigma_\nu \right)$, $L_S$ is the spin e.m. field interaction Lagrangian density [28] and $\sigma_\mu$ is the four-vector spin. $L_{F.P}$ is the Lagrangian density function of Fermi pressure [29],

$$L_{F.P} = \psi^* \left( i\partial_t + \frac{\nabla^2}{2m} + \mu_{ch} \right) \psi + C_0 \psi^* \psi, \tag{2}$$

where, $\psi$ is the spin -1/2 fermionic field and $\mu_{ch}$ is the chemical potential. $L_{B.P}$ is the Bohm-Potential Lagrangian density function [30],

$$L_{B.P} = \frac{\hbar^2}{2m} |\nabla \psi|^2 - e\varphi |\psi|^2 - \frac{\hbar^2}{2m} \frac{\nabla^2 \psi^2}{|\psi|^2} \tag{3}$$

where, $\hbar$ is the Planck's constant and $\varphi$ is the electrostatic potential. We first delineate a procedure for deriving equation of motion of the system through the application of the Euler-Lagrange equations, we get,

$$m\frac{du^\alpha}{d\tau} + \frac{d\Psi}{dt} + \frac{d}{dt}\left[ E - \frac{e^2}{90\pi^2 m^4}(B^2 - E^2)E + \frac{7e^2}{180\pi^2 m^4}(E \cdot B^2) \right] = e^{\alpha\beta} + \frac{\partial}{\partial x_\alpha}(\alpha \sigma_\mu F^{\mu\nu} \sigma_\nu) +$$

$$i\left( \frac{\nabla^2 \Psi}{2m} + \mu\Psi + C_0 \Psi \right) + \nabla \times \left( B - \frac{e^2}{90\pi^2 m^4}\{B^2 - E^2\}E + \frac{7e^2}{180\pi^2 m^4}\{E \cdot B^2\} \right), \tag{4}$$

and

$$\frac{d\sigma^\alpha}{d\tau} + \nabla \left[ \frac{1}{m} \nabla \Psi - \frac{1}{2m} \frac{\nabla \Psi^*}{|\Psi|^2} \right] = \alpha F^{\alpha\mu} \sigma_\nu + \nabla \left[ E - \frac{e^2}{90\pi^2 m^4}(B^2 - E^2)E + \frac{7e^2}{180\pi^2 m^4}(E \cdot B^2) \right] + 2ea|\Psi|. \tag{5}$$

Equations (4) and (5) are the equations of motion for a single electron (using $\hbar = 1$ and $c = 1$). Taking into the consideration $\frac{d}{d\tau} = \gamma \frac{d}{dt} = \gamma \left( \frac{\partial}{\partial t} + u \cdot \nabla \right)$, equations (4) and (5) can be re-written as,

$$m\gamma u^\beta \partial_\beta u^\alpha + \partial_\beta \Psi + u^\alpha k\Psi + \frac{\partial E}{\partial t} - \frac{e^2}{90\pi^2 m^4}\left\{ \left[ 2E \cdot B \frac{\partial B}{\partial t} + B^2 \frac{\partial E}{\partial t} \right] - 3E^2 \frac{\partial E}{\partial t} \right\} + \frac{7e^2}{180\pi^2 m^4}\left[ 2E \cdot B \frac{\partial B}{\partial t} + B^2 \frac{\partial E}{\partial t} \right] =$$

$$eF^{\alpha\beta} u_\beta + \frac{\partial}{\partial x_\alpha} \alpha F^{\mu\nu} \sigma_\nu + \frac{ik^2 \Psi}{2m} + i(\mu + C_0)\Psi + \nabla \times \left[ B - \frac{e^2}{90\pi^2 m^4}(B^2 - E^2)E + \frac{7e^2}{180\pi^2 m^4}(E \cdot B^2) \right], \tag{6}$$

and

$$u^\beta \partial_\beta \sigma^\alpha + \frac{k^2 \Psi}{m} - \frac{k^2 \Psi^*}{2m|\Psi|^2} = \alpha F^{\alpha\nu} \sigma_\nu + \nabla \left[ E - \frac{e^2}{90\pi^2 m^4}(B^2 - E^2)E + \frac{7e^2}{180\pi^2 m^4}(E \cdot B^2) \right] + 2ea|\Psi| -$$

$$\frac{k}{2m|\Psi|^4} + \left[ |\Psi|^3 - 2\left( |\Psi|^2 \nabla \Psi + \Psi^2 \nabla \Psi^* \right) \right]. \tag{7}$$

Averaging over $N$-particle states and following the method from the referenced work [5] on the Pauli's equation, we assign a probability to a Dirac particle being in state α. Applying the ensemble average for any tensorial quantity, the four-velocity is $U^\mu = \langle u^\alpha \rangle$ and the total four-spin density is $S^\mu = \langle \sigma^\mu \rangle$. The microscopic four dimensional velocity and spin density in the rest frame are also defined $v^\mu = u^\mu - U^\mu$ and $\Sigma^\mu = \sigma^\mu - S^\mu$, satisfying $\langle v^\mu \rangle = 0$ and $\langle \Sigma^\mu \rangle = 0$ respectively. Moreover, if the creation of electron pairs are taken, then the conservation of electron number gives

$$\partial_\mu (nU^\mu) = 0, \tag{8}$$

and the macroscopic Maxwell's equation

$$\partial_\mu F^{\mu\upsilon} = -enU^\upsilon. \tag{9}$$

Taking the ensemble average of equations (6) and (7), we obtain the equations

$$U^\beta \partial_\beta U^\alpha + \frac{\partial_\beta \Psi}{m\gamma} + \frac{U^\alpha k\Psi}{m} + \frac{1}{m\gamma}\frac{\partial E}{\partial t} - \frac{e^2}{90\pi^2 m^5 \gamma}\left[\left\{2EB\frac{\partial B}{\partial t} + B^2 \frac{\partial E}{\partial t}\right\} - 3E^2 \frac{\partial E}{\partial t}\right] + \frac{7e^2}{180\pi^2 m^5 \gamma}\left[2EB\frac{\partial B}{\partial t} + B^2 \frac{\partial E}{\partial t}\right] =$$

$$\frac{e}{m\gamma} F^{\alpha\beta} U_\beta + \frac{e}{m^2\gamma}\left\langle \frac{\partial}{\partial x_\alpha}\left(\Sigma_\mu F^{\mu\upsilon}\Sigma_\upsilon\right)\right\rangle + \frac{e}{m^2\gamma}\frac{\partial}{\partial x_\alpha}\left(S_\mu F^{\mu\upsilon} S_\upsilon\right) - \left\langle v^\beta \partial_\beta v^\alpha \right\rangle - U^\beta \left\langle \partial_\beta v^\alpha \right\rangle + \frac{ik^2 \Psi}{2m^2\gamma} +$$

$$\frac{i(\mu + C_0)\Psi}{m\gamma} + \frac{1}{m\gamma}\nabla \times \left[B - \frac{e^2}{90\pi^2 m^4}(B^2 - E^2)E + \frac{7e^2}{180\pi^2 m^4}(B \cdot E)\right], \tag{10}$$

and

$$U^\beta \partial_\beta S^\alpha = \frac{e}{m} F^{\alpha\upsilon} S_\upsilon - \left\langle v^\beta \partial_\beta \Sigma^\alpha \right\rangle - U^\beta \left\langle \partial_\beta \Sigma^\alpha \right\rangle + kE - \frac{e^2 k}{90\pi^2 m^4}(B^2 - E^2)E + \frac{7e^2 k}{180\pi^2 m^4}(E \cdot B^2) +$$

$$2ea\Psi - \frac{k^2}{2m\Psi}\left(\frac{1}{\Psi} - |\Psi|^2 + 3\right). \tag{11}$$

Equations (8), (9), (10), and (11) together provide a comprehensive description of covariant relativistic spin plasma in the context of many particle systems, representing the primary novel contribution of this letter. It is evident that the plasma exhibits a highly linear nature, with terms implicitly summing over all quantum states. The equations reveal a strong coupling between the four velocity and the four-spin density, as these quantities appear intertwined within the formulations. The equations are evidently covariant and will be helpful to put it in the standard vector notation, so the macroscopic four velocities will be written as

$$U^\mu = [\gamma_f, \gamma_f U] \tag{12}$$

where $U$, is the three vector spatial component of velocity and $\gamma_f = U^0 = \langle \upsilon_\alpha^0 \rangle$ is the average relativistic factor associated with the $N$-particle. Using the constraint and the definition for $\upsilon^\mu$ [9], one readily obtains $U_\mu U^\mu = 1 - \langle \upsilon_\mu \upsilon^\mu \rangle$, which when combined with equation (12), yields

$$\gamma_f = \gamma \left(1 - \langle \upsilon_\mu \upsilon^\mu \rangle\right)^{\frac{1}{2}}, \qquad (13)$$

where, $\gamma = \left(1 - U^2\right)^{-\frac{1}{2}}$. There are two aspects for $N$-particle, one from particle in motion in e. m. field and second from thermal motion [31]. We adopt the same technique to formulate the macroscopic vector four spin densities $S^\mu = \left(S^0, S\right)$. Taking into the consideration the covariant constraints, $U^\mu S_\mu = 0$ and the time component can be obtained as, $S^0 = U \cdot S$. Thus, equations (10) and (11) can be written in the vectorial form,

$$\frac{d}{dt}(m\gamma_f) = \frac{e}{\gamma_f}(U \cdot E) + \frac{e}{m\gamma_f^2}\frac{\partial}{\partial t}\Psi_s + \frac{P_c}{\gamma_f} + \frac{e^2}{90\pi^2 m^4}\left[\left\{2B \cdot E \frac{\partial B}{\partial t} + B^2 \frac{\partial E}{\partial t}\right\} - 3E^2 \frac{\partial E}{\partial t}\right] + \frac{7e^2}{180\pi^2 m^4 \gamma_f^2}$$
$$\left[2E \cdot B \frac{\partial B}{\partial t} + B^2 \frac{\partial E}{\partial t}\right] + \frac{ik^2 \psi_s}{2m\gamma_f^2} - \frac{k\psi_s}{\gamma_f} - \frac{1}{\gamma_f}\frac{\partial E}{\partial t} + \frac{i}{\gamma_f^2} ik (\mu + C_0) + \frac{1}{\gamma_f} \nabla \times \left[B - \frac{e^2}{90\pi^2 m^4}\left(B^2 - E^2\right)E + \frac{7e^2}{180\pi^2 m^4}\left(B^2 \cdot E\right)\right]$$
$$-\frac{1}{\gamma_f^2}\frac{d\psi_s}{dt}, \qquad (14)$$

and

$$\frac{d}{dt}(m\gamma_f U) = \frac{e}{\gamma_f}(E + U \times B) + \frac{e}{m\gamma_f^2}\nabla \Psi_s + \frac{F_c}{\gamma_f} - \frac{1}{\gamma_f^2}\frac{d\Psi}{dt} - \frac{k\Psi_s}{\gamma_f} - \frac{1}{\gamma_f^2}\frac{\partial E}{\partial t} + \frac{e^2}{90\pi^2 m^4 \gamma_f^2}\left[\left\{2BE \frac{\partial B}{\partial t} + B^2 \frac{\partial E}{\partial t}\right\} - 3E^2 \frac{\partial E}{\partial t}\right]$$
$$-\frac{7e^2}{180\pi^2 m^4 \gamma_f^2}\left[2EB \frac{\partial B}{\partial t} + B^2 \frac{\partial E}{\partial t}\right] + \frac{ik^2 \Psi}{2m\gamma_f^2} + \frac{i(\mu + C_0)}{\gamma_f} + \frac{1}{\gamma_f^2} \nabla \times \left[B - \frac{e^2}{90\pi^2 m^4}\left(B^2 - E^2\right) + \frac{7e^2}{180\pi^2 m^4}\left(B^2 \cdot E\right)\right] - \frac{\nabla \Pi}{\gamma_f}, \qquad (15)$$

where,

$$\Psi_s = S_\mu F^{\mu\upsilon} S_\upsilon,$$
$$\langle \upsilon^\beta \partial_\beta \upsilon^\alpha \rangle = \langle \partial_\beta (\upsilon^\beta \upsilon^\alpha) - (\partial_\beta \upsilon^\beta) \upsilon^\alpha \rangle,$$
$$P_c = \frac{e}{m\gamma}\frac{\partial}{\partial x_\alpha}\langle \Sigma_\mu F^{\mu\upsilon} \Sigma_\upsilon \rangle - m\langle \partial_\beta (\upsilon^\beta \upsilon^\alpha) - (\partial_\beta \upsilon^\beta) \upsilon^\alpha \rangle - mU^\beta \langle \partial_\beta \upsilon^\beta \rangle,$$

and

$$F_c = \frac{e}{m\gamma_f}\left\langle \frac{\partial}{\partial x_\alpha}\left(\Sigma_\mu F^{\mu\upsilon} \Sigma_\upsilon\right) \right\rangle - m\langle \partial_\beta (\upsilon^\beta \upsilon^\alpha) - (\partial_\beta \upsilon^\beta) \upsilon^\alpha \rangle - mU^\beta \langle \partial_\beta \upsilon^\alpha \rangle.$$

Equations (14) and (15) are the time and spatial components of equation (10) respectively. The electrostatic spin potential is represented as $\psi_s$ and $\Pi$ is the relativistic quantum pressure tensor. Additionally, $P_c$ and $F_c$ account for the nonlinear corrections to the power and force arising from the microscopic four dimensional velocity and spin density in the rest frame of fluid. The orders of equation (11) can be written as

$$\frac{dS^0}{dt} = \frac{e}{m\gamma_f} E \cdot S + \frac{P_s}{\gamma_f} + \frac{kE}{\gamma_f} - \frac{e^2 k}{90\pi^2 m^4 \gamma_f}(B^2 - E^2)E + \frac{7e^2 k}{180\pi^2 m^4 \gamma_f}(E \cdot B^2) + \frac{2ea\psi}{\gamma_f} - \frac{k^2 \psi}{2m\gamma_f}\left[\frac{1}{|\psi|} - |\psi|^2 + 3\right],$$ (16)

and

$$\frac{dS}{dt} = \frac{e}{m\gamma_f}\left[S^0 E + S \times B\right] + \frac{\Xi_s}{\gamma_f} - \frac{kE}{\gamma_f} - \frac{e^2 k}{90\pi^2 m^4 \gamma_f}\left[(B^2 - E^2)E\right] + \frac{7e^2}{180\pi^2 m^4 \gamma_f}(B^2 E) + \frac{2ea\psi}{\gamma_f} - \frac{k^2 \psi}{2m\gamma_f}\left[\frac{1}{|\psi|} - |\psi|^2 + 3\right],$$ (17)

where,

$$\langle \upsilon^\beta \partial_\beta \Sigma^\alpha \rangle = \langle \partial_\beta(\upsilon^\beta \Sigma^\alpha) - (\partial_\beta \upsilon^\beta)\Sigma^\alpha \rangle,$$

$$P_s = -\langle \partial_\beta(\upsilon^\beta \Sigma^\alpha) - (\partial_\beta \upsilon^\beta)\Sigma^\alpha \rangle - U^\beta \langle \partial_\beta \Sigma^\alpha \rangle,$$

and

$$\Xi_s = -\langle \partial_\beta(\upsilon^\beta \Sigma^\alpha) + (\partial_\beta \upsilon^\beta)\Sigma^\alpha \rangle - U^\beta \langle \partial_\beta \Sigma^\alpha \rangle.$$

Equations (16) and (17) are the time and spatial components of equation (11). $K$ is the relativistic quantum thermal- spin coupling tensor. $P_s$ and $\Xi_s$ are the nonlinear corrections to the spin evolution caused by the microscopic four dimensional velocity and spin density in the rest frame of fluid. Equations (14), (15), (16), (17) and the equation of continuity form a full set of macroscopic quantum electrodynamical equations. These equations describe the conceptual framework and computational efficiency of this formalism, emphasizing the interaction between electromagnetic waves and plasma while incorporating relativistic effects. This formalism governs interaction dynamics utilizing critical parameters such as, the relativistic correction factor $\gamma_f$, the temporal component of the four-spin vector $S^0$, velocity field coupling, the Bohm potential, and Fermi pressure. The derived quantum electrodynamic (QED) equations provide a framework for investigating relativistic and quantum phenomena across various physical regimes, including astrophysical plasmas, laser plasma interactions, quantum devices, and extreme field environments. This framework enhances the understanding of spin dynamics and collective behaviour within plasma systems. Future research will aim to refine theoretical models and improve experimental methodologies to broaden the applicability of these findings.

**Discussion** - In this letter, we have developed a novel model for relativistic quantum spin-1/2 plasmas using a covariant Lagrangian formulation rooted in QED. The developed model would be applicable to all energy regimes. Considering the Lagrangian densities for free electron, electron - e. m. wave interaction and Fermi pressure in the relativistic quantum plasma regime, our model describes a wave nature of electrons and spin. Due to high electron velocities, they show relativistic effects and the spin effects in the equations accounts the intrinsic angular momentum of electron and its interaction with e. m. field. These results in the phenomena like spin orbit coupling and spin dependent forces, which affect the trajectory and plasma dynamics in the strong e. m. field. The Bohm potential in the equation is useful for understanding quantum tunneling and interference effect in plasma. The Lagrangian density formulation for the Heisenberg - Euler is the one loop effective action correction term in QED that describes the non-linear interaction of e. m. field in vacuum. It incorporates the first order correction due to vacuum polarization which is the creation and annihilation of virtual electron - positron pairs in the strong e. m. field. By considering these terms, the model leads to phenomena like photon - photon scattering. This QED based model provides a more accurate description of quantum plasma behaviour than existing mildly relativistic or non-relativistic quantum models. It incorporates spin effects and relativistic dynamics, crucial for understanding high density plasmas. This work will open new insights in fusion science and contribute in the development of future high-density plasma experiments, table-top accelerators, compact gamma photon sources, high-gain free electron lasers, collective atomic recoil lasers and quantum plasmonic devices. The model offers a comprehensive framework for exploring relativistic quantum plasma dynamics across all energy regimes, including collective spin effects with applications in astrophysics and laser-plasma interactions.

## ACKNOWLEDGEMENTS

The author thanks SERB- DST, Govt. of India for financial support under MATRICS scheme (grant no: MTR/2021/00471

## DATA AVAILABILITY

The data that support the findings of this study are available from the corresponding author upon reasonable request.